\documentclass[letterpaper,12pt]{article}   
\usepackage{osajnl2} 
\usepackage[draft]{hyperref} 
\usepackage{amsfonts}
\usepackage{amssymb}
\usepackage{amsmath}
\usepackage{graphicx}
\usepackage{txfonts}

\begin{document}

\title{Quantum discord induced by white noises}


\author{Jia-sen Jin, Chang-shui Yu,$^{*}$ Pei Pei, and He-shan Song}

\address{School of Physics and Optoelectronic Technology,\\
Dalian University of Technology, Dalian 116024 China}
\address{$^*$Corresponding author: ycs@dlut.edu.cn}

\begin{abstract}We discuss the creation of quantum discord between two two-level
atoms trapped in an optical cavity in a noisy environment. It is
shown that nonzero steady-state quantum discord between atoms can be
obtained when the white-noise field is separately imposed on atoms
or cavity mode, while the steady-state quantum discord reaches zero
if both cavity mode and atoms are driven simultaneously by
white-noise fields. In particular, we demonstrate that white-noise
field in different cases can play a variously constructive role in
the generation of quantum discord.
\end{abstract}


\maketitle 

\section{Introduction}

Quantum entanglement, as a valuable physical resource, has been
widely applied to most quantum-information processing tasks [1-3].
It is a type of correlation lies in the superposition principle and
the tensor product structure of the state space of composite quantum
systems. However, entanglement can not distinguish the classical
correlation and quantum correlation encoded in a quantum system
exactly, since some other kinds of nonclassical correlations could
exist even if the entanglement is vanishing. In order to capture the
pure quantum correlations in a quantum system, Ollivier and Zurek
\cite{OZ} introduced the so-called quantum discord, which is the
discrepancy between two quantum extensions of classical mutual
information. Similar idea was suggested by Henderson and Vedral
[5,6] simultaneously and independently. Quantum discord is a more
general measure of quantum correlation may include entanglement but
is an independent one [7,8], it is more useful than entanglement to
describe the quantum correlation involved in a quantum system.
Moreover, it is shown that quantum discord can provide a speedup for
some certain tasks over their classical counterparts both
theoretically [9-12] and experimentally \cite{exp}, in this sense
quantum discord can also be regarded as a physical resource in
quantum information theory, so it is desirable to investigate the
quantum discord further and more broadly.

On the other hand, a real quantum system will interact with the
environmental noise inevitably. The interaction between the system
of interest and the noise, which usually models surroundings of the
system, leads to a decoherence process [14,15]. As a consequence,
the system may end up in a mixed state which is no longer suitable
for quantum information processing. In order to prevent or minimize
the environmental noise, numerous proposals have been made, such as
loop control strategies [16,17], quantum error correction [18,19]
and exploiting the decoherence-free subspace [20-22]. Recent studies
show that quantum discord is more robust than entanglement against
decoherence. For example, under the conditions in which entanglement
sudden death (ESD) \cite{Yu} can occur, the quantum discord will
just vanish asymptotically [24-26] or instantaneously at some time
points \cite{Wang}. Moreover, quantum discord can not be destroyed,
even can be frozen, under some certain decoherence channels [28,29].
In spite of this, this kind of quantum correlation will still be
destroyed due to the destructive effects of the environmental noise
at the asymptotic infinite-time limit.

In this paper, we investigate the generation of quantum discord in a
noisy environment, by which we find that environmental noise can
also play a constructive role in creation of quantum discord. Our
system consists of two two-level atoms trapped in a leaky optical
cavity. Both the cavity mode and the atoms are driven by external
white-noise fields. We focus on the problem of generating quantum
discord with only incoherent sources are available. The results show
that the quantum discord of the two atoms can indeed be created and
even reach a stable value in such a situation. This paper is
organized as follows. In Sec.II, we firstly introduce our model and
recall the calculations of quantum discord briefly, and then discuss
the noise-assisted generation of quantum discord detailedly and
compared with the generation of entanglement in three different
cases: only atoms are driven by white-noise field; only cavity mode
are driven by white-noise field; and both cavity mode and atoms are
driven by the same white-noise field. The conclusion is drawn
finally.

\section{Quantum discord induced by white noises}

\begin{figure}
\includegraphics[width=1\columnwidth]{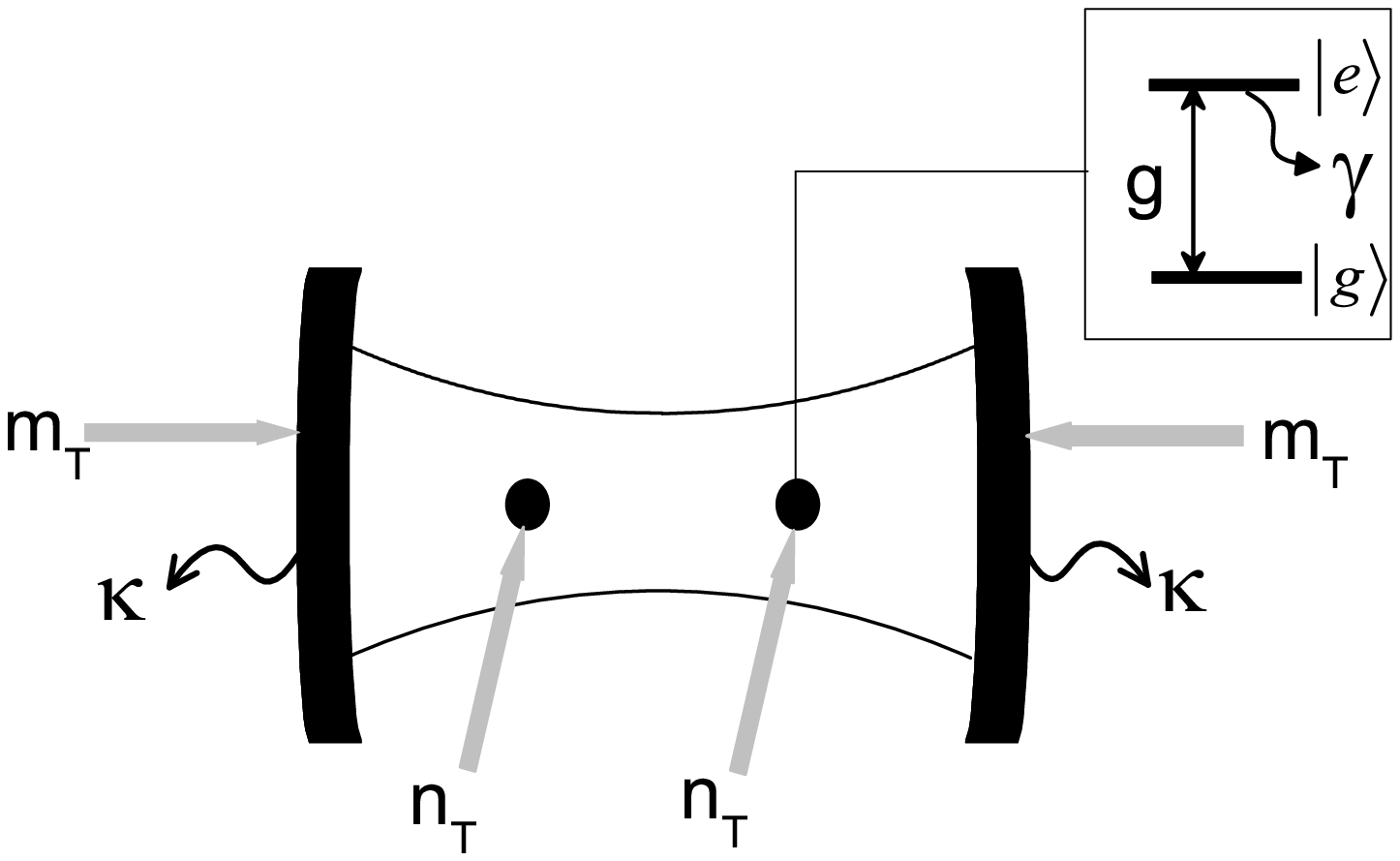}
\caption{} \label{1}
\end{figure}

Our system consists of two identical two-level atoms, labeled by 1
and 2, trapped in a leaky optical cavity, see Fig. \ref{1}. Each
atom has a ground state $|g\rangle$ and an excited state
$|e\rangle$. The coupling strength between atoms and cavity mode is
$g$. We assume that the distance between the atoms is much larger
than an optical wavelength, therefor dipole-dipole interaction can
be neglected. The cavity mode and atoms are driven by the external
white-noise fields, intensities of which will be characterized in
terms of effective photon numbers $m_T$ and $n_T$, respectively. The
Hamiltonian [30-32] describing the interaction between atoms and
cavity mode reads
\begin{equation}
H=\omega_ca^\dagger
a+\sum_{i=1,2}\omega_a\sigma_i^+\sigma_i^-+g(a^\dagger
\sigma_i^-+\mathrm{H.c.}),
\end{equation}
where $\omega_a$ is the transition frequency of atom, $\omega_c$ is
the frequency of the cavity mode, $a$ is the annihilation operator
of cavity mode, and $\sigma_i^-$ ($i=1,2$) is the lower operator of
the $i$th atom. To simplify the representation, we turn to the
interaction picture with respect to $H_0=\omega_ca^\dagger
a+\sum_{i=1,2}\omega_a\sigma_i^+\sigma_i^-$, now the Hamiltonian is
given by
\begin{equation}
H_I=\sum_{i=1,2}g(a^\dagger\sigma_i^-+\mathrm{H.c.}),
\end{equation}
where we have assumed the atom-cavity coupling is on resonance.

There are two channels that the relaxation of atom-cavity system can
take place, one is atom spontaneous emission at rate $2\gamma$ and
the other is cavity decay at rate $2\kappa$. The master equation
governing the time evolution of the global system is given by
(setting $\hbar=1$)
\begin{equation}
\dot{\rho}=-i[H,\rho]+\mathcal {L}(\rho),
\end{equation}
with the Liouvillean $\mathcal {L}(\rho)$ as follow
\begin{eqnarray}
\mathcal
{L}(\rho)&=&(n_T+1)\gamma\sum_{i=1,2}(2\sigma_i^-\rho\sigma_i^+-\sigma_i^+\sigma_i^-\rho-\rho\sigma_i^+\sigma_i^-)\cr\cr&&+n_T\gamma\sum_{i=1,2}(2\sigma_i^+\rho\sigma_i^--\sigma_i^-\sigma_i^+\rho-\rho\sigma_i^-\sigma_i^+)\cr\cr&&+(m_T+1)\kappa(2a\rho
a^\dagger-a^\dagger a\rho - \rho a^\dagger
a)\cr\cr&&+m_T\kappa(2a^\dagger\rho a-aa^\dagger\rho-\rho
aa^\dagger).
\end{eqnarray}
Here we do not specify the white-noises, but their intensities $m_T$
and $n_T$ refer to the effective particle numbers.

Before our discussion, it is necessary to recall the quantum discord
briefly. For a bipartite quantum system, if $\rho^{ab}$ denotes the
density operator of the joint system, and $\rho^{a}$ ($\rho^{b}$)
denotes the reduced density operator of subsystem $a$ ($b$), then
the quantum discord between subsystems $a$ and $b$ can be obtained
as follow
\begin{equation}
\mathcal {Q}(\rho)=\mathcal {I}(\rho)-\mathcal {C}(\rho),
\end{equation}
where $\mathcal {I}(\rho)=S(\rho^a)+S(\rho^b)-S(\rho^{ab})$ is the
quantum mutual information and $\mathcal {C}(\rho)$ is the classical
correlation between the two subsystems. As discussed in ref. [5,6],
the classical correlation is provided by $\mathcal
{C}(\rho)=\mathrm{max}_{\{B_k\}}[S(\rho^a)-S(\rho|\{B_k\})]$, where
$\{B_k\}$ is a set of von Neumann measurements performed on
subsystem $b$ locally, $S(\rho|\{B_k\})=\sum_{k}p_kS(\rho_k)$ is the
quantum conditional entropy, $\rho_k=(\mathbb{I}\otimes
B_k)\rho(\mathbb{I}\otimes B_k)/\mathrm{Tr}(\mathbb{I}\otimes
B_k)\rho(\mathbb{I}\otimes B_k)$ is the conditional density operator
corresponding to the outcome labeled by $k$, and
$p_k=\mathrm{Tr}(\mathbb{I}\otimes B_k)\rho(\mathbb{I}\otimes B_k)$.
Here $\mathbb{I}$ is the identity operator performed on subsystem
$a$.

Due to the complicated optimization involved, it is usually
intractable to obtain the analytical expressions of quantum discord.
So far as we know, the analytical expression of quantum discord is
obtained only for some specific states [7,8], hence we will
calculate the quantum discord of the atoms in a numerical way.
Generally, the maximization of the classical correlations is
achieved by a positive operator-valued measure (POVM) [5,6,33],
however, for two qubits system, which is our case, Hamieh \emph{et
al.} \cite{Hamieh} have proved that the projective measurement is
the POVM, which maximizes the classical correlations. Therefore, it
is sufficient for us to evaluate the discord using the following set
of projectors,
$\{|\psi_1\rangle\langle\psi_1|,|\psi_2\rangle\langle\psi_2|\}$, in
which
$|\psi_1\rangle=\cos{\theta}|g\rangle+e^{i\phi}\sin{\theta}|e\rangle$
and
$|\psi_2\rangle=-\cos{\theta}|e\rangle+e^{-i\phi}\sin{\theta}|g\rangle$.
For the purposes for obtaining the maximum of
$S(\rho^a)-S(\rho|\{B_k\})$, the parameters $\theta$ and $\phi$ vary
from $0$ to $2\pi$. We cutoff the intracavity photon numbers at a
value of 5 in the simulations. In this paper, we will only focus on
the cases of only incoherent sources are available, \emph{i.e.}, the
initial state of the joint system is
$|g\rangle_1|g\rangle_2|0\rangle_c$ where the subscript 1 or 2
denotes the atom 1 or 2 and subscript 'c' denotes cavity mode. In
order to explicitly show the dependence of quantum discord on the
noise intensities, atom spontaneous emission rate and cavity leaky
rate, we will examine the noise-induced quantum discord in the
following three different cases: noise drives only the cavity mode;
noises drive the two two-level atoms only; and noises drive both
cavity mode and atoms.

\begin{figure}
\includegraphics[width=1\columnwidth]{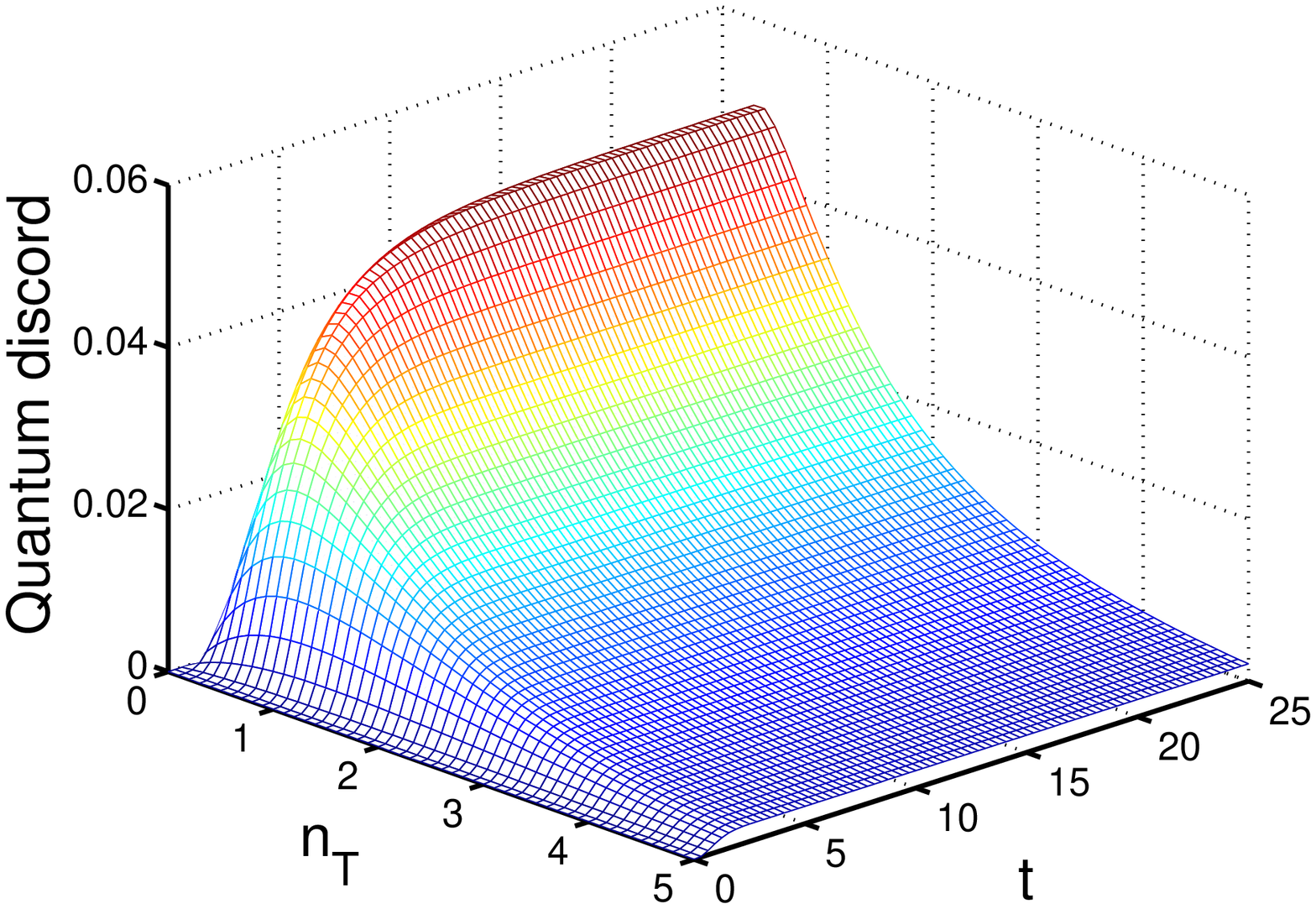}
\caption{} \label{2}
\end{figure}

Firstly, we will examine the noise-induced quantum discord between
atoms in the case of only atoms are driven by noises, thereby we can
set $m_T=0$ in Eq. (4). We have plotted the amounts of both quantum
discord of the two two-level atoms as a two-variable function of the
intensity of the noise $n_T$ and time $t$ in Fig. \ref{2}. For
$n_T=0$, we can see that there is no quantum correlation between two
atoms at any time, namely the vacuum field can not induce quantum
discord between atoms. For a finite value of $n_T$, it is shown that
steady-state quantum discord between two atoms can be generated. The
behavior of the amount of steady-state quantum discord is
nonmonotonic with noise intensity, it increases to a maximum value
for an optimal intensity of the noise and then decreases. The
maximum value of steady-state quantum discord appears at a small
noise intensity ($n_T\approx0.7$). The larger $n_T$ is not conducive
to construct steady-state quantum discord. It is especially
interesting that for a given noise intensity the quantum discord
between atoms evolves from zero to a stationary value monotonically,
which is quite different from the case of generation of entanglement
where entanglement increases first and then drops [30-32]. That is
to say the white-noise always plays a constructive role for quantum
discord between atoms during the time evolution.

\begin{figure}
\includegraphics[width=1\columnwidth]{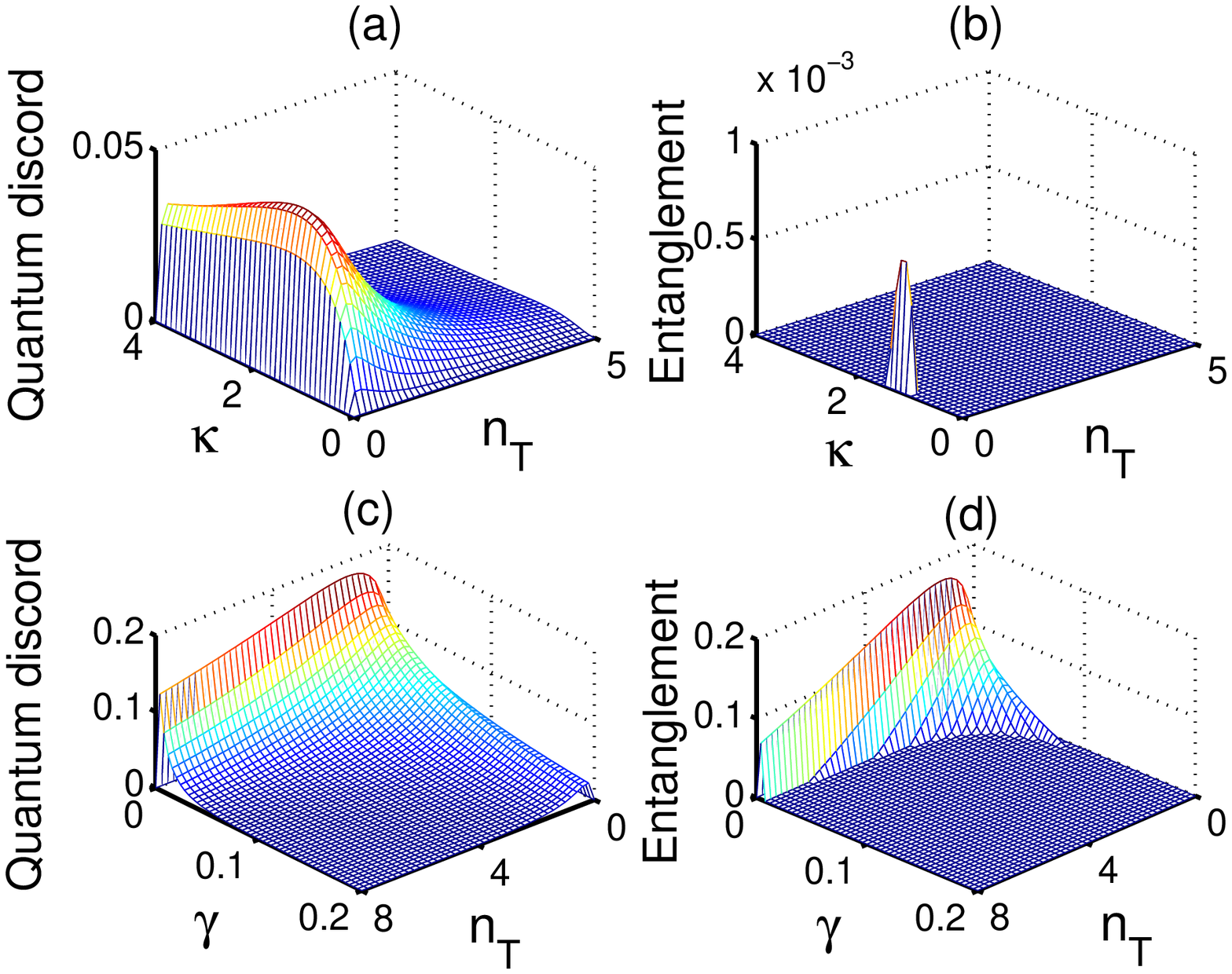}
\caption{ } \label{3}
\end{figure}

It is also worthwhile to study the steady-state quantum discord of
the atoms as a function of intensity of noise and the cavity leaky
rate or the atom spontaneous emission rate, these results are shown
in Fig. \ref{3} (a) and (c), respectively. As a comparison, we also
show the steady-state entanglement between atoms in Fig.\ref{3} (b)
and (d). From the upper two plots in Fig. \ref{3}, one can find that
a proper cavity leaky rate can help increase the steady-state
quantum discord of the two atoms, but the cavity leakage almost has
no effects on creation of steady-state entanglement, the
steady-state entanglement only rises in a very small region and is
extremely tiny. From Fig.\ref{3}(c), one can find that the behavior
of the steady-state quantum discord between atoms exhibits a double
resonance on both atomic spontaneous emission rate and noise
intensity at an intermediate value. Contrasting to the steady-state
quantum discord, which decreases to zero asymptotically with $n_T$
and $\gamma$ increasing, the steady-state entanglement between atoms
vanishes suddenly and drastically, see Fig. \ref{3}(d), this
phenomenon is reminiscent of the well known ESD \cite{Yu}, where the
entanglement of a system will suddenly disappears when interacting
with a noisy environment. This comparison between Fig.\ref{3}(c) and
(d) shows that the quantum discord is more robust than entanglement
against the noisy environments.

\begin{figure}
\includegraphics[width=1\columnwidth]{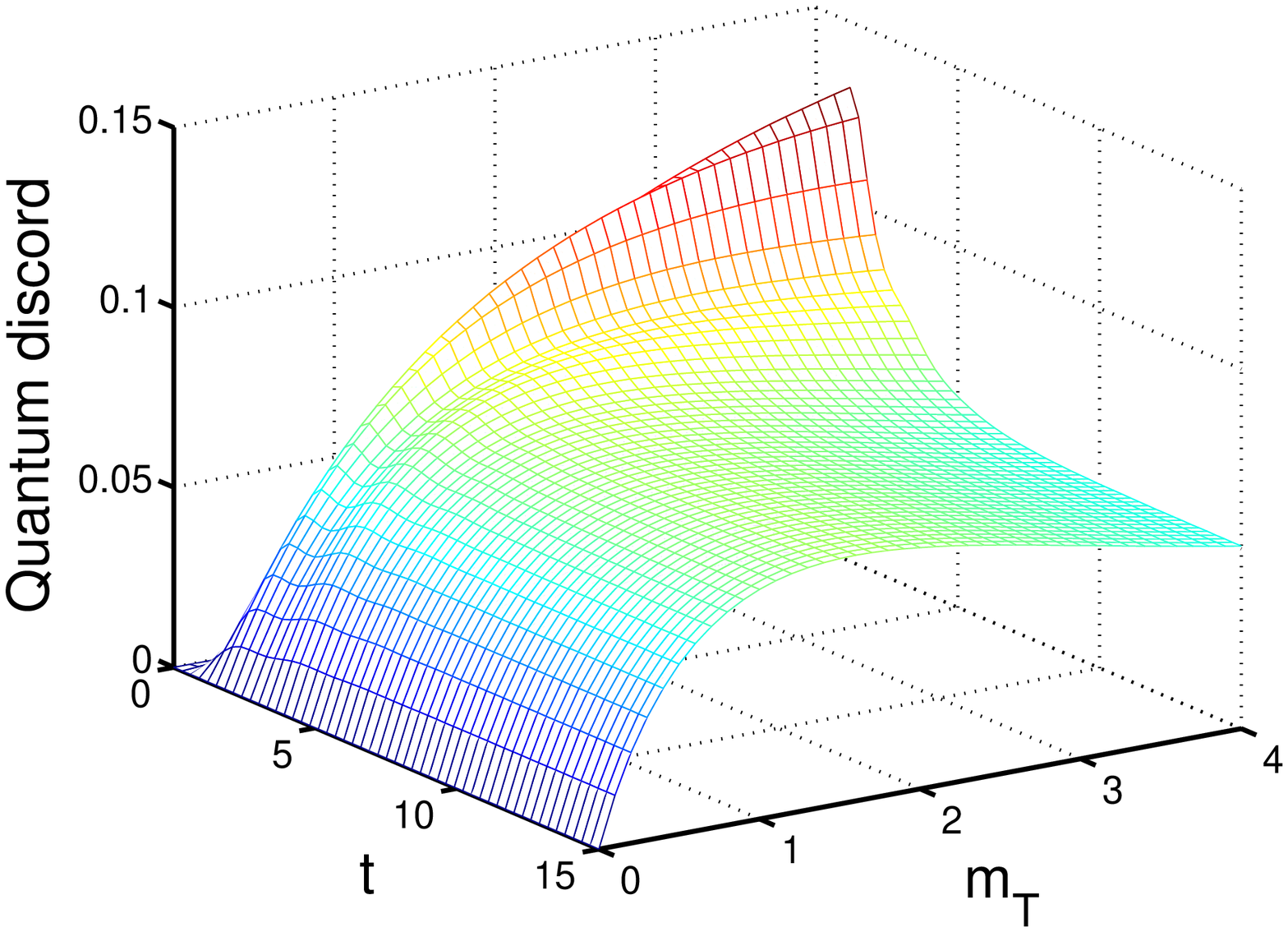}
\caption{} \label{4}
\end{figure}

Now we come to the case that noise is only imposed on the cavity
mode, \emph{i.e.}, $n_T=0$. The time evolutions of quantum discord
and between atoms is shown in Fig. \ref{4}. Similar to Fig. \ref{2},
the quantum discord between atoms evolves to a stationary value for
a nonzero $m_T$. The steady-state quantum discord increases to a
maximum value for an optimal noise intensity. Contrast to Fig.
\ref{2}, the steady-state quantum discord holds a relative larger
value when the noise intensity is large. That is to say, when noise
is imposed on the cavity mode only, it is more easier to obtain the
quantum discord between atoms than that the noises are imposed on
the atoms. Moreover, we can see from Fig. \ref{4} that the behavior
of quantum discord between atoms is nonmonotonic during the time
evolution for a given noise intensity, it firstly increases to a
maximum value and then decreases to a stationary value. It means
that the noise intensity plays both constructive and destructive
roles in the generation of quantum discord between atoms. At the
beginning the constructive effect is dominate, later the destructive
effect is dominate and finally they are balanced where the quantum
discord arrives at a stationary value. A weak vibration during the
evolution can also be found in a short time especially obviously for
$m_T\approx1$. These are somewhat like the case for entanglement.

\begin{figure}
\includegraphics[width=1\columnwidth]{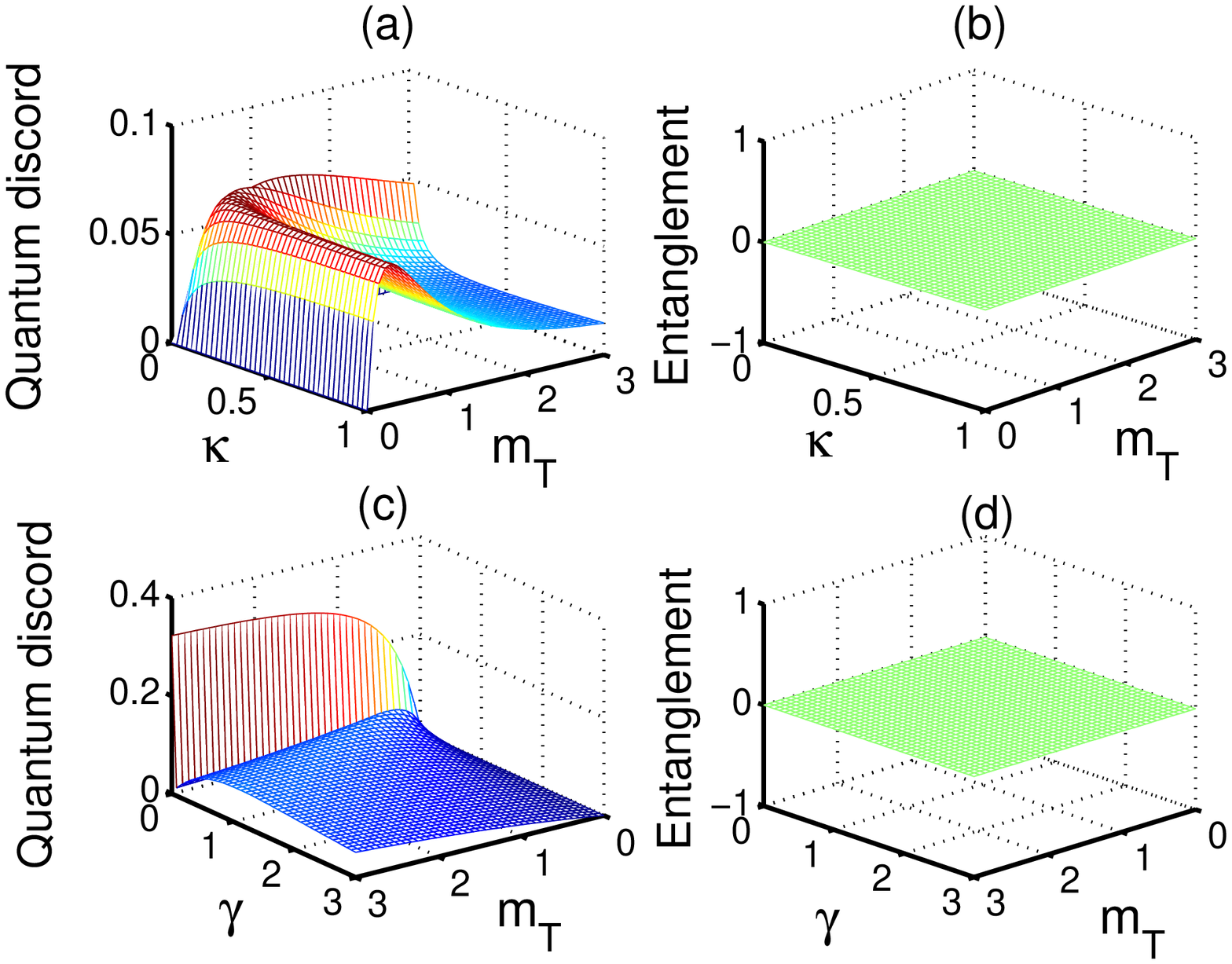}
\caption{} \label{5}
\end{figure}

The dependence of the steady-state quantum discord and entanglement
between atoms on the cavity leaky rate and atom spontaneous emission
rate are shown in Fig. \ref{5}. From the right two plots in Fig.
\ref{5}, one can find that, the steady-state entanglement between
atoms can not be generated irrespective of the value of $\kappa$ and
$\gamma$. From Fig. \ref{5}(a) one can see that the behavior of
steady-state quantum discord between atoms shows a resonance on both
cavity leaky rate and noise intensity like Fig. \ref{3}(c). From
Fig. \ref{5}(c), it is surprising that the steady-state quantum
discord decreases rapidly along with increasing of atom spontaneous
emission and then a slight revival occurs with the atomic
spontaneous emission rate increasing for a larger $m_T$. This means
that an appropriate atomic spontaneous emission is conducive for
generating quantum discord between atoms if the environmental noise
is strong.

\begin{figure}
\includegraphics[width=1\columnwidth]{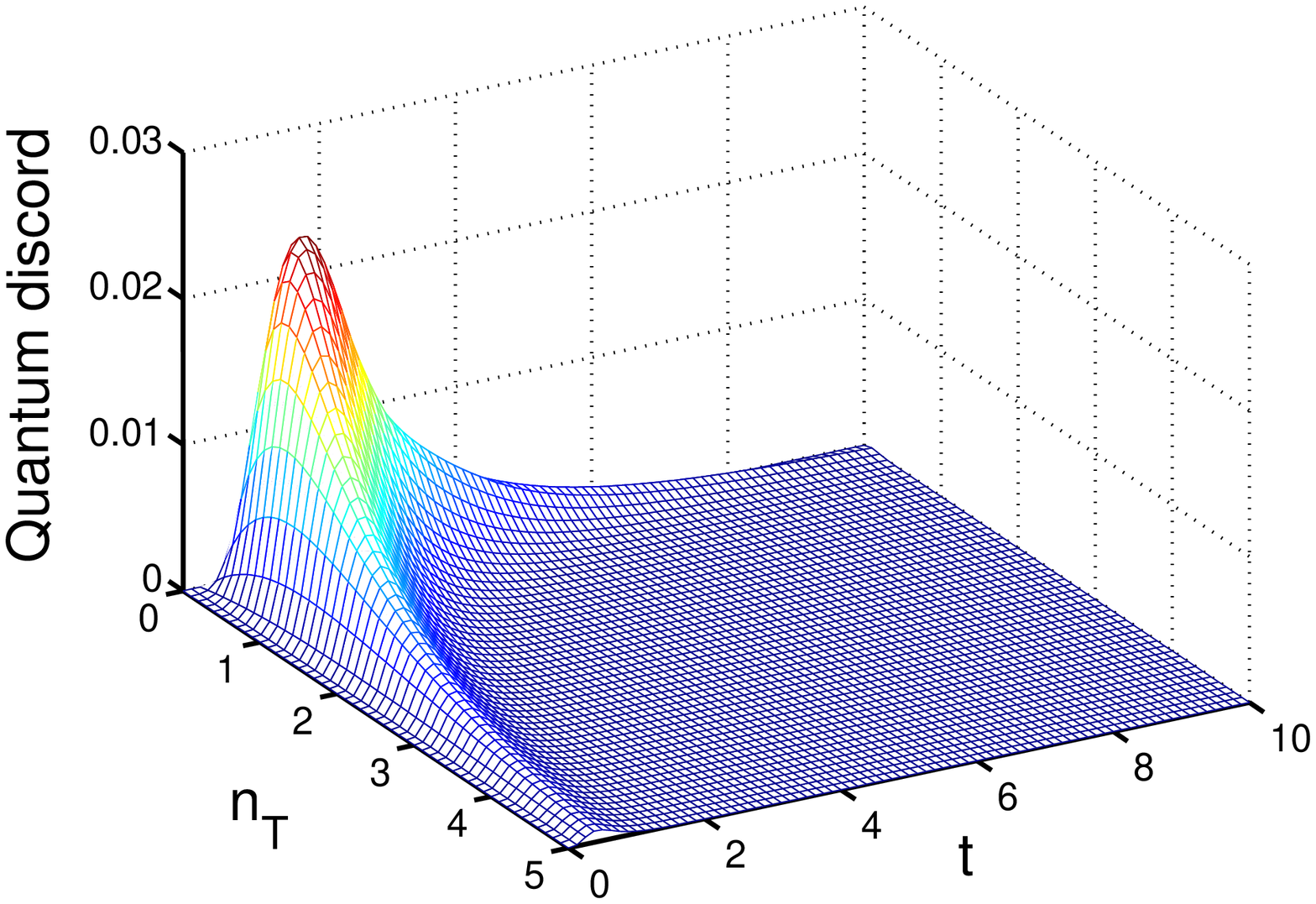}
\caption{} \label{6}
\end{figure}

In the following, we will consider that noises are imposed on both
cavity mode and atoms. Here we assume the two noises are the same,
\emph{i.e.}, $n_T=m_T$. The quantum discord between atoms as a
function of noise intensity ($n_T=m_T$) and time are shown in Fig.
\ref{6}. Again the entanglement between atoms can not be generated.
However, different from the previous two cases, when cavity mode and
atoms are driven by the same noise, the quantum discord between
atoms evolves to zero instead of a stationary nonzero value for any
noise intensity. The quantum discord between atoms during the time
evolution is nonmonotonic about noise intensity, it increases to a
maximal value for an optimal noise intensity and then
 asymptotically decreases to zero for a large intensity. This can be
understood as that the combination of the noises imposed on the
cavity modes and atoms strengthens the destructive role in quantum
discord, even though they have different effects on the quantum
discord separately. The joint effects of the external noises lead to
a vanishing quantum discord between atoms.

In an experimental scenario, our atomic level structures can be
achieved by alkali metal atoms and the leaky optical cavity can be
realized by either the conventional Fabry-Perot cavity or the new
type microtoroidal cavity [35-37]. The parameters concerning in this
scheme is the coupling constant $g$, cavity leaky rate $\kappa$ and
atomic spontaneous emission rate $\gamma$. In the current available
experiments [38,39], they take $g\thicksim2\pi\times100\mathrm{MHz}$
and $g^2\thicksim20\gamma\kappa$. With these parameters, the quantum
discord will come to a steady value at about $1\mu s$.

\section{Conclusion and discussion}

In conclusion, we have discussed the creation of quantum discord
between two two-level atoms trapped in a leaky optical cavity in
three different cases. It has been shown that there exist quite
differences between the noise-assisted generation of quantum discord
and quantum entanglement. One can  find that nonzero steady-state
quantum discord between atoms can be generated when the white-noise
field is separately imposed on atoms or cavity mode, even though the
steady-state quantum discord reaches zero when both cavity mode and
atoms are driven by two identical white-noise fields. Although
quantum discord between atoms in the noisy environment is small but
it can be distillable \cite{distill}, which may provide a broader
way to acquire quantum information resource and to process quantum
computation. In addition, the effects of cavity leaky rate and atom
spontaneous emission rate on the quantum discord between atoms are
also discussed. These results may play a guiding part in generating
desirable quantum discord between atoms in noisy environment.

\section{Acknowledgements}

One of the authors (J. S. J.) thanks Dr. X. L. Huang for valuable
discussions. This work was supported by the National Natural Science
Foundation of China, under Grant No. 10805007 and No. 10875020, and
the Doctoral Startup Foundation of Liaoning Province.

\newpage
Fig.1 Schematic illustration of the system composed of an optical
cavity and two two-level atoms. The cavity leaky rate is $2\kappa$
and the cavity mode is driven by the white-noise field with
intensity $m_T$. The atom has a ground state $|g\rangle$ and an
excited state $|e\rangle$ with atom spontaneous emission rate
$2\gamma$ (see the inset), and the atoms are driven by the
white-noise field with intensity $n_T$.

Fig.2 (Color online) Quantum discord of the two atoms between atoms
as a function of noise intensity $n_T$ and time $t$ (in units of
$1/g$) in the case of $m_T=0$. The parameters are chosen as
$\gamma=0.1g$ and $\kappa=1.5g$.

Fig.3 (Color online) (a)The steady-state quantum discord and (b) the
steady-state entanglement between atoms as a function of noise
intensity $n_T$ and cavity leaky rate $\kappa$ (in units of $g$) in
the case of $m_T=0$ and $\gamma=0.1g$. (c)The steady-state quantum
discord and (d) the steady-state entanglement between atoms as a
function of $n_T$ and atom spontaneous emission rate $\gamma$ (in
units of $g$) in the case of $m_T=0$ and $\kappa=2g$.

Fig.4 (Color online) Quantum discord between atoms between atoms as
a function of noise intensity $m_T$ and time $t$ (in units of $1/g$)
in the case of $n_T=0$. The parameters are chosen as $\kappa=0.1g$
and $\gamma=0.2g$.

Fig.5 (Color online) (a) The steady-state quantum discord and (b)
the steady-state entanglement versus noise intensity $m_T$ and
cavity leaky rate $\kappa$ (in units of $g$) in the case of $n_T=0$
and $\gamma=0.1g$. (c) The steady-state quantum discord and (d) the
steady-state entanglement versus $m_T$ and atom spontaneous emission
rate $\gamma$ (in units of $g$) in the case of $n_T=0$ and
$\kappa=0.1g$.

Fig.6 (Color online) Quantum discord between atoms between atoms as
a function of noise intensity $n_T(=m_T)$ and time $t$ (in units of
$1/g$). The parameter are chosen as $\kappa=g$ and $\gamma=0.1g$.

\newpage

\newpage

\newpage

\newpage

\newpage

\end{document}